\documentstyle[12pt]{article}

\def\ltsim{\lower3pt\hbox{$\, \buildrel < \over \sim \, $}}
\def\gtsim{\lower3pt\hbox{$\, \buildrel > \over \sim \, $}}
\def\be{\begin{equation}}
\def\ee{\end{equation}}
\def\ba{\begin{eqnarray}}
\def\ea{\end{eqnarray}}
\def\bea{\begin{eqnarray}}
\def\eea{\end{eqnarray}}

\def\ga{\mathrel{\raise.3ex\hbox{$>$\kern-.75em\lower1ex\hbox{$\sim$}}}}
\def\la{\mathrel{\raise.3ex\hbox{$<$\kern-.75em\lower1ex\hbox{$\sim$}}}}

\begin{document}

\baselineskip=16pt
\begin{titlepage}
\rightline{IHES/P/02/52}
\rightline{LPT-Orsay-02/55}
\rightline{OUTP-02/08P}
\rightline{HIP-2002-34/TH}
\rightline{hep-th/0208036}
\rightline{August  2002}
\begin{center}

\vspace{1cm}

\large {\bf  DBI ACTION FROM CLOSED STRINGS AND D-BRANE
 SECOND QUANTIZATION}

\vspace{1 cm}
\normalsize

\centerline{
Ian I. Kogan\footnote{i.kogan@physics.ox.ac.uk}$^{\,a,b,c}$  and
 Dimitri Polyakov\footnote{ polyakov@pcu.helsinki.fi}$^{\,d}$
}
\smallskip
\medskip
$^a$
{\em IHES, 35 route de Chartres, 91440, Bures-sur-Yvette,  France }\\
$^b$
{\em Laboratoire de Physique Th\'eorique,
Universit\'e de Paris XI, 91405 Orsay C\'edex, France
}\\
$^c$
{\em Theoretical Physics, Department of Physics, Oxford University,
 1 Keble Road, Oxford, OX1 3NP,  UK}\\
$^d$ {\em Department of Physical Sciences, University of Helsinki
and Helsinki Institute of Physics, PL 64, FIN-00014 Helsinki, Finland}
\smallskip

\vskip0.6in \end{center}

\centerline{\large\bf Abstract}
Brane-like vertex operators play an important role in a world-sheet
formulation
 of D-branes and M theory. In this paper we derive the DBI
D-brane
action  from closed NSR   string  with brane-like states.
 We also show that  these operators carry RR charges and
define D-brane wave functions in a second quantized formalism.

\vspace*{2mm}

\end{titlepage}

\section{Introduction}
One of the most challenging problems in  string theory is to find a way to
 quantize background and, as a part of this big problem, how to describe
 second quantization of different branes. In a canonical world-sheet
formulation branes are not on the same footing as perturbative string
states,
  which are produced by vertex operators (quantum objects) in open and
closed
sectors  and are second-quantized. To find  quantum operator(s) which can
be
interpreted  as brane(s)  creation (annihilation) operator(s) will be very
important. Some time ago two- and five-forms brane-like vertex operators
in NSR superstring theory
 were introduced in \cite{Polyakov:1997uy} and in a recent
 paper \cite{Polyakov:2001zr}  detailed proof has been given that these
operators are physical, i.e.  BRST-invariant
and BRST-nontrivial. While these  vertex operators  are proven to be
physical,
clearly there are no massless two-forms or five-forms in
perturbative spectrum of an open string, so the question arises
what is the actual role of these physical states in superstring theory.
It has been suggested in the  papers \cite{Polyakov:2000xc},
 \cite{Kogan:2000nw}  that these vertex operators
and the closed string brane-like vertices which can be constructed from
the
open ones  carry  crucial information about nonperturbative physics of
strings, D-branes and M-theory, rather than being related to perturbative
string dynamics.
The first hint comes from the superalgebraic arguments as the zero
momentum parts of these operators  appear as two and fiveform central
terms in a picture changed space-time superalgebra \cite{Polyakov:1997uy}
Since the p-form central terms in the SUSY algebra are always
related to topological charges of p-branes \cite{Townsend}
this gave  the first indication that these new  states  may be related to
brane dynamics.
The essential property of these operators is the ghost-matter mixing
  i.e. they appear only as   higher BRST-cohomologies
breaking the discrete symmetry of the picture-changing
\cite{Friedan:1985ge}.
 Using the formalism of brane-like states
one can study the dynamics of branes and
M-theory  \cite{Kogan:2000nw} as well as of superstring theory
in curved backgrounds such as
$AdS_5\times{S^5}$ \cite{Polyakov:2000xc}.  The fact that one
has a matter-ghost mixing in the presence of these operators leads to the
 the world-sheet  logarithmic CFT \cite{Kogan:2000nw},\cite{Kogan:2001ku}
 which agrees with earlier  suggestions that string theory with
second-quantized   D-branes (and other backgrounds in general) must
 be described by  world-sheet  logarithmic CFT \cite{Kogan:1995df} where
 logarithmic operators describe background  collective coordinates. It
seems
 quite natural to assume that by this reason brane-like vertex operators
 must describe D-brane collective coordinates.

 In this paper we shall further analyse these  brane-like vertex operators
 and find several properties which strongly support the idea to use tham
  as creation-annihilation
 operators for extended solitonic objects (D-branes). To get it we must
 see that our vertex operators have  exactly the same number of
 physically independent polarizations as the number of transverse
directions,
 otherwise we shall not be able to describe the correct number of D-brane
 collective coordinates. At the same time we have to show that   there is
a correct coupling to the RR field. In this paper we shall
 only discuss the case of $D3$ brane and respective vertices.
 We shall show
 that close string brane-like vertex corresponding to the $D3$ brane
 has precisely $6$ physical degress of freedom which  is what we need for
 $D3$ brane. But it is  open
string brane-like vertex operator which has the desired coupling to the RR
 field.
 From the low energy effective action point of view, we will show that
the close string brane-like vertices generate the bulk
(square root of the determinant) terms in the DBI action, while open
string brane-like
vertices correspond to the RR terms.
 The fact that one must   have  two  type of vertex operators to
describe second- quantized solitonic object in string theory is rather
  unusual and  deserves further   investigation.  We also want to
stress here  that the brane operators are not usual
creation-annihilation operators in a sense that adding more brane
operators  does not lead to creation of additional branes but rather
 generates  interaction for brane collective coordinates.

The paper is organized as follows.
In the next  section we shall review basic facts about open
and closed string brane-like states. In a section 3
we will derive the DBI effective action
straight from the sigma-model with closed string brane-like states
by computing appropriate correlation functions on the sphere.
It is remarkable that the effective action will be shown to
have the $1/g_{st}$ dilaton dependence
even though technically the computations are performed
in closed string theory on the sphere. As we shall argue the
``open-string''
  dilaton dependence follows from the logarithmic nature of the
world-sheet
CFT when the pair of closed string brane-like operators produce a
logarithmic
 cut effectively producing disk from the  sphere.
 We shall be also able to reproduce the correct D-brane tension from
correlators of the closed string brane-like states.
In a section 4, we compute the disc correlators of open string
vertices  with the Ramond Ramond insertion on the disc with Neumann
boundary condition showing that
the open string brane operators  are  carriers of the RR-charges.
In conclusion we discuss obtained results and  unsolved problems which
 have to be addressed in a future.

\section{Brane-like states}

Brane-like states  are described
by physical vertex operators, existing at
selected nonzero pictures only, i.e. the vertices with
 superconformal ghost matter mixing which cannot be removed
by a picture changing transformation. Unlike usual perturbative
vertex operators, such as a graviton or a photon,
the brane-like states do not account for any point-like string excitation
but describe the dynamics of nonperturbative extenmded objects
in string theory, such as branes and solitons.
These brane-like vertex operators appear in both open and closed
superstring theories. In the open string sector these vertices, the
two-form and the five-form are given by:

\bea
\label{open}
V_5^{(-3)}(k)
&=&H_{m_1...m_5}(k)
\oint{dz\over{2i\pi}}e^{-3\phi}\psi_{m_1}...\psi_{m_5}e^{ikX}(z)
\nonumber \\
V_5^{(+1)}(k)
&=& H_{m_1...m_5}(k)
\oint{dz\over{2i\pi}}(e^{\phi}\psi_{m_1}...\psi_{m_5}
+B_{m_1...m_5}+C_{m_1...m_5})e^{ikX}(z) \nonumber \\
V_2^{(-2)}(k)
&=&H_{m_1m_2}(k)
\oint{dz\over{2i\pi}}e^{-2\phi}\psi_{m_1}\psi_{m_2}e^{ikX}(z) \\
V_2^{(0)}(k)
&=&H_{m_1m_2}(k)
\oint{dz\over{2i\pi}}(\psi_{m_1}\psi_{m_2}+C_{m_1m_2})
e^{ikX}(z)
\nonumber
\eea
Here $X^m$, $m=0,...9$ are 10d space-time coordinates, $\psi_m$
are their superpartners onn the worldsheet, $\phi$
and $\chi$ are bosonized
superconformal ghost fields.
BRST-invariance and BRST nontriviality of the states (1)
has been proven in \cite{Polyakov:2001zr}.
The open string five-form state can be expressed at pictures -3 and +1
(but with no version at picture zero). The picture $+1$ version
must also include the  b-c ghost counterterms
$B_{m_1...m_5}$ and $C_{m_1...m_5}$
(carrying the reparametrization
ghost numbers $-1$ and $+1$) in order to insure its BRST-invariance;
the picture $-3$-version requires no b-c ghost terms.
The precise form of these ghost counterterms has been given in
\cite{Polyakov:2001zr}. In this paper we will only need the expression
for $B_{m_1...m_5}$:

\bea
B_{m_1...m_5}=
{1\over{10}}
{({\hat{cT_\chi}})_7}
(\partial\phi-\partial\chi)\partial{b}be^{2\phi-\chi}
\psi_{m_1}...\psi_{m_5}
(\psi\partial^2{X})
\eea
The hat operators
$({\hat{cT_\chi}})_n$ are defined by acting on any local
operator $A(w)$ as
\bea
({\hat{cT_\chi}})_nA(w) &=&\lim_{u\rightarrow{w}}
\oint{dz\over{2i\pi}}(z-u)^n:cT:(z)A(u)
\nonumber \\
T_\chi &=& {1\over2}((\partial\chi)^2-\partial^2\chi)
\eea

Analogously, the BRST invariant
expression for the two-form vertex at picture zero must
contain the c-ghost two-form $C_{m_1m_2}$.
BRST nontriviality of the vertices (1) can be proven straightforwardly
by showing that they have non-vanishing correlation functions with other
physical vertex operators, however the BRST non-triviality imposes
important constraints on the five-form $H_{m_1...m_5}$
Indeed, it is possible to construct the operator
\bea
W_5=H_{m_1...m_5}\oint{{dz}\over{2i\pi}}e^{\chi-4\phi}\partial\chi
\psi_{m_1}...\psi_{m_5}(\psi_m\partial{X^m}){e^{ikX}}
\eea
such that
\bea
\lbrace{Q_{BRST}},W_5\rbrace=-{3\over2}
H_{m_1...m_5}\oint{{dz}\over{2i\pi}}e^{-3\phi}
\psi_{m_1}...\psi_{m_5}{e^{ikX}}
\nonumber \\
+iH_{m_1...m_5}\oint{{dz}\over{2i\pi}}
\partial^2ce^{\chi-4\phi}\partial\chi
(k\psi)\psi_{m_1}..\psi_{m_5}e^{ikX}
\nonumber \\
\equiv V_5^{(-3)}(k)
+iH_{m_1...m_5}\oint{{dz}\over{2i\pi}}
\partial^2ce^{\chi-4\phi}\partial\chi
(k\psi)\psi_{m_1}..\psi_{m_5}e^{ikX}
\eea
Therefore the $V_{5}^{(-3)}$ vertex operator
is BRST-nontrivial only if the second term in the commutator
is non-vanishing. This is obviously equivalent to the condition
\be
k_{{\lbrack}m_6}H_{m_1...m_5\rbrack}\neq{0}
\ee
or
\be
dH^{(5)}\neq{0}
\ee
which means that the
$V_5$-operator is physical if the five-form $H(k)$ is not closed.
In fact, this condition has a simple physical meaning:
below  we will see that $dH$ plays the role  of the D-brane wavefunction
in the second-quantized formalism which of course must not vanish.
Analogous condition can be shown to appear for the five-form
at the $+1$-picture, though in that case the form of the
$W_5$-operator is much more complicated. We will not present it here for
the sake of shortness.

Apart from the open string vertices (\ref{open})
  brane-like states are also present
in the closed string sector. They can be constructed straightforwardly by
either taking the brane-like  forms (1) as holomomorphic and
antiholomorphic
parts of the vertices, or
taking  the brane-like  left parts and photonic right parts.
Depending on the contraction of the  space-time indices,
the closed string brane-like states may describe various
branes and brane configurations.
In this paper we shall explore the closed string
brane-like states given by

\bea
V_5^{cl.(-3)}=H_{m_1...m_6}(k)\int{d^2z}
e^{-3\phi-\bar\phi}\psi_{{m_1}}...\psi_{m_5}\bar\psi_{m_6}
e^{ikX}
\nonumber \\
V_5^{cl.(+1)}
=H_{m_1...m_6}(k)\int{d^2z}
(e^{-3\phi}\psi_{{m_1}}...\psi_{m_5}
+B_{{}m_1...m_5}+C_{{}m_1...m_5})
e^{-{\bar\phi}}\bar\psi_{m_6}
e^{ikX}
\nonumber \\
+c.c.
\eea
where $H_{m_1...m_6}$ is antisymmetric 6-form field.
Complex congugated part must be added here to insure the overall unitarity
of the amplitudes.
In this paper we restrict ourselves
to considering only a totally antisymmetric rank 6 representation
of the Lorentz group, as it is this representation that will yield
the desired D3-brane dynamics.
The meaning of other possible irreducible rank 6 representations
 in (8) is of course an interesting question and we hope to consider
it in the future. It is possible that these representations
correspond to more complicated brane configurations
(such as brane  bound states) but in any case we do not discuss
any of these questions in our  present paper.
Below we will show that the physical meaning of the 6-form
is that it determines the space-time location of the D3-brane,
after imposing appropriate BRST constraints on H.

\section{DBI action from closed string brane like states}

In this section we derive the DBI action for the D3-brane from the
closed string sigma-model with the 6-form brane-like states (8).
To obtain the low energy effective action
one first of all has to specify the number of physical degrees of freedom
associated with the 6-forms $H_{m_1...m_6}(k)$  left after fixing
the gauge symmetry associated with the BRST conditions on $H$.

Consider the operator $V_5^{cl.(-3)}$.
As for the BRST- nontriviality of its left part,
it still requires the condition
\be
k_{{\lbrack}m_7}H_{m_1...{m_5\rbrack}m_6}\neq{0}
\ee
where the brackets imply total antisymmetrization.
The condition (9) is easily derived in full analogy with (5).
Indeed, in the closed string case the role of the $W_5$-operator
is played by
$$
W_5=H_{m_1...m_5}\int{d^2z}e^{\chi-4\phi-{\bar\phi}}\partial\chi
\psi_{m_1}...\psi_{m_5}{\bar\psi_{m_6}}
(\psi_m\partial{X^m}){e^{ikX}}
$$
and
$$
\lbrace{Q_{BRST}},W_5\rbrace=
 V_5^{(-3)cl.}(k)
+{i\over2}H_{m_1...m_6}\oint{{dz}\over{2i\pi}}
{\partial^2}ce^{\chi-4\phi-{\bar\phi}}\partial\chi
(k_n\psi_n)\psi_{m_1}..\psi_{m_5}{\bar\psi_{m_6}}e^{ikX}
$$
Again, the BRST nontriviality means that the second term does not vanish.

At the same time, the condition for the BRST-invariance of the
antiholomorphic part leads
to the
constraint
\be
k_{m_1}H_{m_1...m_6}(k)={div}H^{(6)}={0}
\ee
The constraints (9), (10)  have  simple geometrical meaning.
They imply that for any given polarization
$m_1,...m_6$ of the $V_5$ vertex , its momentum
must be  transverse to the
$m_1,...m_6$ directions so the vertex effectively
propagates in 4 dimensions for any given polarizations.
The BRST constraints (9), (10) on H largely reduce the number of
independent physical degrees of freedom.
To identify the degrees of freedom, note that the number of independent
components
of a general 6-form in 10 dimensions is equal to
${{10!}\over{4!6!}}=210$.
The BRST condition (10) means that the 6-form $H^{(6)}$ can be locally
expressed as  divergence of the 7-form which reduces the number of
independent components to
\be
N={{10!}\over{3!7!}}-{{10!}\over{2!8!}}+{{10!}\over{1!9!}}-1
=84
\ee
This number is still reduced by constraints induced by
the nontriviality conditions (9).
To calculate it note that the  conditions (9) induce the set of
the gauge transformations
\be
H^{(6)}\rightarrow{H^{(6)}}+d\Lambda^{(5)}
\ee
The number of gauge transformations is given by
\be
P={{10!}\over{5!5!}}-{{10!}\over{4!6!}}+{{10!}\over{3!7!}}
-{{10!}\over{2!8!}}
+{{10!}\over{1!9!}}-1
=126
\ee
However, what we need is not the full set of these gauge transformations
but only those consistent with the constraints (9).
This implies that we need to include only the gauge transformations
that can be written  in terms of the Laplacian of the 6-forms.
The number of transformations to exclude is then given by
the number of seven-forms that are closed and not divergences of
the eight-forms; it is easy to check that
this number is given by
\bea
Q=2{\times}{{10!}\over{4!6!}}-{{10!}\over{5!5!}}
-{{10!}\over{3!7!}}=48
\eea
Therefore the total number of independent gauge transformations
is equal to $126-48=78$
 number of physical degrees of freedom left
after imposing the BRST constraints
is given by
\be
N-P+Q=84-78=6
\ee

Below we shall see that they correspond to 6 transverse
fluctuations of a D3-brane.
Having identified the number of the degrees of freedom,
we can choose the gauge
\be
V_5^{(-3)}=\epsilon_{{\lbrack}t_1...t_5}\lambda_{t_6\rbrack}(k^\perp)
\psi_{t_1}...\psi_{t_5}\bar\psi_{t_6}e^{i{k^\perp}X}+c.c.
\ee

where the space-time indices are split in the $4+6$ way :
$m=(a,t), a=0,...3;t=4,...9$ and ${k^\perp}X=k_aX_a$.
Note that the 4+6 splitting is achieved as a result of a particular
choice of the gauge (16) fixing the 6 physical degrees of freedom.
Indeed this gauge has a SO(1,3) x SO (6) isometry.
Choosing the gauge (16) is equivalent here to fixing one particular
polarization of the vertex (8), for which the momentum
is orthogonal to 6 space-time indices.
Therefore the vertex (8) effectively propagates in four-dimensional
space-time. This situation is quite different, for example, from the case
of photons, for which the transversality condition does not
reduce the effective dimensionality in which they propagate.
Indeed, for the gauge choice (16) only one polarization is admissible,
while while in the photonic case there is nothing fixing polarization.
Technically, the difference between the photonic and the brane-like case
occurs because a single porarization condition for photons
$(ke(k))=0$ is replaced by two BRST conditions (9), (10)
which altogether are much stronger than transversality
constraints for standard operators of U(1) gauge fields.

Now we are ready to begin the computation of
the low energy effective action.
It is easy to see that the three-point function of the $V_5$-operators
iz zero, as the three-point correlator of the NSR fermions vanishes:
$$<:\psi_{t_1}...\psi_{t_5}:(z_1)
:\psi_{s_1}...\psi_{s_5}:(z_2)
\psi_{u_1}...\psi_{u_5}:(z_3)>=0$$.
Therefore the first nonvanishing contribution to the beta-function
and low-energy equations of motion comes from
the 4-point correlator of the $V_5$-vertices.
We will consider the limit of a slowly changing
$\lambda^t$-field in which only the massless poles
of the Veneziano amplitude are important.
We will also need a picture-changed version of the
$V_5^{(-3)}$-operator to insure correct ghost number balance
in the holomorphic part of the 4-point correlator.
Acting on $V_5^{(-3)}$ with picture changing transformation we get the
picture $-2$-representation of the 5-form:
\bea
V_5^{(-2)}=H_{m_1...m_5}(k)\oint{{dz}\over{2i\pi}}
c\partial\chi{e^{\chi-3\phi}}\partial\chi
\psi_{m_1}...\psi_{m_5}e^{ikX}(z)+c.c.
\eea
Analogous picture-changing transformation
can be done for the left (brane-like) part of the closed-string 6-form.
To compute the four-point function
of the $V_5^{closed}$ , one has to take
the left (5-form) part of two operators at picture
-2 while the left part of the remaining two at picture $+1$.
Then the contribution to the four-point function will be given by
\bea
A_4(p,k,q,l)=
<c\partial\chi{e^{\chi-3\phi}}\partial\chi
\psi_{t_1}...\psi_{t_5}(\bar\partial{X_{m_6}}
i(p^\perp\bar\psi)\bar\psi_{t_6}
e^{ip^\perp{X}}(z_1,\bar{z}_1) \nonumber \\
c\partial\chi{e^{\chi-3\phi}}\partial\chi
\psi_{s_1}...\psi_{s_5}(\bar\partial{X_{s_6}}
i(k^\perp\bar\psi)\bar\psi_{s_6}
e^{ik^\perp{X}}(z_2,\bar{z}_2)
\nonumber \\
e^{-\bar\phi}B_{u_1...u_5}\bar\psi_{u_6}e^{i{q^\perp}X}
(z_3,\bar{z}_3)
e^{-\bar\phi}B_{v_1...v_5}\bar\psi_{v_6}e^{i{l^\perp}X}
(z_4,\bar{z}_4)>
\nonumber \\
\times\epsilon_{\lbrack t_1...t_5}\lambda_{t_6\rbrack}(p^\perp)
\epsilon_{\lbrack s_1...s_5}\lambda_{s_6\rbrack}(k^\perp)
\epsilon_{{\lbrack}u_1...u_5}\lambda_{u_6\rbrack}(q^\perp)
\epsilon_{{\lbrack}v_1...v_5}\lambda_{v_6\rbrack}(l^\perp)>+c.c.
\eea
Here the indices $s,t,u,v=4,...9$ and
the B-fiveform, carrying the fermionic ghost number -1,
has been defined above in (2), (3).

For simplicity let us first  consider the
case when only one out of six components
of $\lambda_t$ is nonzero;
for instance one can take
$$\lambda_4\equiv\lambda;\lambda_{5,6,7,8,9}=0$$.
It will then be straightforward
 to generalize it to the case when all the components are nonzero.
First, let us calculate the antiholomorphic photonic part of this
correlator
(which of course is holomorphic in the  complex conjugated part).
Simple calculation gives
\bea
A_{R}(p,k,q,l)=
{{1}\over{({\bar{z_1}}-{\bar{z_2}})^2({\bar{z_3}}-{\bar{z_4}})^2}}
-{\lbrace}
{{({k^\perp}{{p}^{\perp}})\delta({p^\perp}+{k^\perp}+{q^\perp}+{l^\perp})}
\over{(\bar{z_1}-\bar{z_2})
(\bar{z_3}-\bar{z_4})}}
({1\over{(\bar{z_1}-\bar{z_2})
(\bar{z_3}-\bar{z_4})}}
\nonumber \\
-{1\over{(\bar{z_1}-\bar{z_3})
(\bar{z_2}-\bar{z_4})}}
+{1\over{(\bar{z_1}-\bar{z_4})
(\bar{z_2}-\bar{z_3})}}){\rbrace}
\eea
Here and elsewhere we only consider the kinematic
part of the amplitude, dropping the factor of
${\prod_{i,j}}|z_i-z_j|^{k_ik_j}$ since we are
only interested in cotributions from the massless poles.
Next, calculation of the holomorphic matter part
gives
\bea
A_L^{matter}(p,k,q,l)=
({1\over{(\bar{z_1}-\bar{z_2})
(\bar{z_3}-\bar{z_4})}}
-{1\over{(\bar{z_1}-\bar{z_3})
(\bar{z_2}-\bar{z_4})}}
+{1\over{(\bar{z_1}-\bar{z_4})}})^5
{1\over{z_3-z_4}}
\nonumber \\
{\times}
({{p_a}\over{(z_1-z_3)^2}}
+{{k_a}\over{(z_2-z_3)^2}}
+{{l_a}\over{(z_3-z_4)^2}})
({{p_a}\over{(z_1-z_4)^2}}+
{{k_a}\over{(z_2-z_4)^2}}+
{{q_a}\over{(z_3-z_4)^2}})
\nonumber \\
{\times}\delta({p^\perp}+{k^\perp}+{q^\perp}+{l^\perp})
\eea
Next, we have to calculate the ghost contribution.
It is given by the correlator
\bea
A_L^{ghost}=
\oint{{du}\over{2i\pi}}
\oint{{dw}\over{2i\pi}}
(u-z_3)^7(w-z_4)^7
<ce^{\chi-3\phi}\partial\chi(z_1)
ce^{\chi-3\phi}\partial\chi(z_2)
\nonumber \\
b\partial{b}e^{2\phi-\chi}(\partial\phi-\partial\chi)(z_3)
b\partial{b}e^{2\phi-\chi}(\partial\phi-\partial\chi)(z_4)
c(\partial\chi\partial\chi-\partial^2\chi)(u)
c(\partial\chi\partial\chi-\partial^2\chi)(w)>
\eea

Technically this part is the most complicated
since it involves evaluating two tedious contour integrals
entering the definition (3) of the hat operators.
Upon calculating these integrals,  the
 final answer is greatly simplified after fixing the Koba-Nielsen's
measure:
\bea
z_1\rightarrow\infty
\nonumber \\
z_2=z
\nonumber \\
z_3=1
\nonumber \\
z_4=0
\eea
multiplying by the $SL(2,C)$ FP determinant
\be
det(SL(2,C))=|z_1-z_3|^2|z_1-z_4|^2|z_3-z_4|^2\sim{|z_1|^4}
\ee
and integrating over z.
After fixing the SL(2,C) gauge,
expression for the antiholomorphic
part of the correlator becomes
\bea
A_R(p,k,q,l)=(p^\perp{k^\perp})
{\delta{(p^\perp+k^\perp+q^\perp+l^\perp)}}
{1\over{{\bar{z_1}}^2}}
(1+{1\over{\bar{z}(\bar{z}-1)}})
\eea
Multiplying it by the holomorphic matter part and the
FP determinant we obtain  the correlator without the
holomorphic ghost factor given by:
\bea
A_R{\times}{A_L^{matter}}\times{det(SL(2,C)}(p,k,q,l)
=z_1^{-3}(1+{1\over{{z}({z}-1)}})^5
(1+{1\over{\bar{z}(\bar{z}-1)}})
\nonumber \\
{\times}(kp){\delta{(p^\perp+k^\perp+q^\perp+l^\perp)}}
({{k_a}\over{(z-1)^2}}+{{l_a}})({{k_a}\over{z^2}}+q_a)
+O(z_1^{-4})
\eea
Evaluating the contour integrals and the ghost holomorphic part, we
obtain
\bea
A_L^{ghost}=
z_1^{3}F(z)+O(z_1^2)
\eea
Since
$A_R{\times}{A_L^{matter}}\times{det(SL(2,C)}(p,k,q,l)$
behaves as $z_1^{-3}$ as we fix the gauge $z_1\rightarrow\infty$,
only the $ z_1^3$ term in the $A_L^{ghost}$ contributes to the correlator.
Evaluation of the $F(z)$ function gives
\bea
F(z)=4{z^3(z-1)^3}\lbrace
({5\over{z(z-1)}}-1)+{1\over{(z-1)^2}}-{1\over{z^2}}
\nonumber \\
+{{z^5}\over{(z-1)^2}}+{{(z-1)^5}\over{z^2}})
({2\over{z-1}}+1)(-{2\over{z}}+1)z^3(z-1)^3
\nonumber \\
+4z^3(z-1)^3\lbrace
(1-z)(-{1\over{(z-1)^2}}+
({1\over{z-1}}-1)(-{{1}\over{z-1}}+5)
+{1\over2}(-{1\over{z-1}}+5)^2
\nonumber \\
-{1\over2}
({1\over{(z-1)^2}}+5))
+z(-{1\over{z^2}}+1-({1\over{z}}+1)(5+{1\over{z}})
+{1\over2}(5+{1\over{z}})^2
-{1\over2}(5+{1\over{z^2}})^2)\rbrace
\nonumber \\
\times
\lbrace{1\over{(z-1)^2}}({2\over{z}}+1)
+{1\over{z^2}}({2\over{z-1}}-1)-
({2\over{z}}-1)({2\over{z-1}}+1)({1\over{z+1}}+{1\over{z}})\rbrace
\nonumber \\
+4z^3(z-1)^3\lbrace({{1-z}\over{z^2}}-{2\over{z}}+{{1-z}\over{z}}-1)
+(9z-3)(-{1\over{z^2}}+({1\over{z}}+{1\over{z-1}})({2\over{z}}
+1))
\nonumber \\
+(9z+6)(-{1\over{(z-1)^2}}+({1\over{z}}+{1\over{z-1}})
({2\over{z}}+1))\rbrace
\nonumber \\
+4z^3(1-z)^3({{1-z}\over{z^2}}-{2\over{z}}+
{{1-z}\over{z}}-1)
\eea
Finally, collecting all the pieces together and integrating over
$z$ gives:

\bea
A(p^\perp,{k^\perp},{q^\perp},{l^\perp})
={\int{d^2z}}{\lbrace}A_R{\times}A_L^{matter}{\times}A_L^{ghost}
{\times}det(SL(2,C))+c.c.{\rbrace}
\nonumber \\
={\lbrace}({k^\perp}{p^\perp})
\lbrack({k^\perp})^2-({p^\perp})^2+({q^\perp}{k^\perp})
+({q^\perp}{p^\perp})\rbrack
+{1\over2}{({k^\perp}{p^\perp})^2}
\nonumber \\
{\times}\lambda(p^\perp))
\lambda({k^\perp})\lambda({q^\perp})\lambda({l^\perp})
\delta({p^\perp}+{k^\perp}+{q^\perp}+{l^\perp}){\rbrace}\Gamma(0)
\eea
where the $\Gamma(0)$ factor accounts for the massless pole
in the $z$  integration.

Using the on-shell momentum conservation
it is convenient to add to this expression the piece given by
\bea
B=-{\lbrace}{1\over4}({k^\perp}s)(p^\perp)^2
-{3\over4}({q^\perp}s)({p^\perp}{ k^\perp})
+{1\over4}({p^\perp}s)({k^\perp}s)
-{1\over8}({p^\perp}{ k^\perp}){s^2}{\rbrace}
\nonumber \\
{\times}
\lambda({p^\perp})\lambda({k^\perp})\lambda({q^\perp})\lambda({l^\perp})
{}
\delta(s)\Gamma(0)
\eea
where
\be
s={p^\perp}+{k^\perp}+{q^\perp}+{l^\perp}
\ee
Adding this piece
corresponds to adding full derivative terms to the
space-time effective action.
Then the expression for the amplitude becomes
\bea
A({p^\perp},{k^\perp},{q^\perp},{l^\perp})
={\lbrace}(p^\perp)^2({k^\perp}{q^\perp})
-({k^\perp}{p^\perp})({p^\perp}{q^\perp})
\lambda({p^\perp})\lambda({k^\perp})\lambda({q^\perp})\lambda({l^\perp})
\rbrace\delta(s)\Gamma(0)
\eea
This concludes the evaluation of the
 the 4-point function (limited to the massless pole contribution).
In the position space, the worldsheet RG equations and the low energy
equations of motion
following from this amplitude are given by

\bea
{{d\lambda}\over{d(log\Lambda)}}=-\partial_a\partial^a\lambda
+\partial_b\partial^b\lambda\partial_a\lambda\partial^a\lambda
-\partial_a\partial_b\lambda\partial^a\lambda\partial^b\lambda=0
\eea
where $a=0,...3$
It is easy to check that this EOM follows from the effective action
given by
\be
S(\lambda)=\int{d^4x}
{\sqrt{det({\eta_{ab}}+\partial_a\lambda\partial_b\lambda)}}
\ee
Generalization of this result for the case when all the six
components
of $\lambda$ are nonzero is completely straightforward,
and all the computations are quite analogous,
though the answer quite predictably follows from
 considerations of the Lorentz invariance.
In the general case, we compute the
amplitude to be
\bea
A({p^\perp}{k^\perp},{q^\perp},{l^\perp})
={\lbrace}({p^\perp})^2({k^\perp}{q^\perp})
+2({k^\perp}{q^\perp})({k^\perp}{p^\perp})
+({p^\perp}{q^\perp})({p^\perp}{k^\perp})
\rbrace
\nonumber \\
{\times}
({\lambda_t({k^\perp})}{\lambda^t({k^\perp})})
({\lambda_s({k^\perp})}{\lambda^s({k^\perp})})
\delta({p^\perp}+{k^\perp}+{q^\perp}+{l^\perp})
\Gamma(0)
\eea
The corresponding low energy equations of motion are given by
\bea
-\partial_b\partial^b\lambda_t
+2\partial_a\partial_b\lambda_s\partial^a\lambda_t\partial^b\lambda^s
\nonumber \\
+\partial_a\partial_b\lambda_t\partial_a\lambda_s\partial^b\lambda^s
+\partial_b\partial^b\lambda_s\partial_a\lambda^s\partial^a\lambda_t=0
\eea
These equations of motion can be easily shown to follow from the quartic
term in the expansion of the DBI action for the D3-brane:
\be
S=\int{d^4{x}}
{\sqrt{det({\eta_{ab}}+\partial_a\lambda_t\partial_b\lambda^t)}}
\ee
This concludes the derivation of the DBI action from sigma-model
with the brane-like states; however, this derivation has been
done in the absense of the dilaton background.
In the next section we will calculate the impact of the dilaton
field on the effective action (36).
\section{Dilaton coupling and brane tension}
In the presence of the dilaton background
the low energy effective Lagrangian (36)
is modified as
$$e^{\alpha\varphi}{\sqrt{det(\eta_{ab}
+\partial_a\lambda_t\partial_b\lambda^t)}}$$.
The problem now is to determine the coefficient $\alpha$.
In closed string theory one usually has $\alpha=-2$ for the dilaton
coupling with the NS-NS fields and $\alpha=0$ for the
coupling with Ramond-Ramond
fields. In our case, however, the situation is different.
To compute the dilaton coupling
one has to consider the three-point function
$<V_5^{closed}V_5^{closed}V_\varphi>$
 This correlator gives the first order term
in the dilaton expansion and which is enough to read off
$\alpha$.
The simple computation gives us
\bea
<V_5^{closed(+1,-1)}(p^\perp,z_1,{\bar{z_1}})
V_5^{closed(-3,+1)}(k^\perp,z_2,{\bar{z_2}})V_\varphi^{0,0}(q^\perp,
z_3,{\bar{z_3}})>
\nonumber \\
=-{{({k^\perp}{q^\perp})\lambda_t({p^\perp})\lambda^t({k^\perp})
\varphi({q^\perp})}\over
{|z_1-z_2|^2|z_1-z_3|^2|z_2-z_3|^2}}\delta(k^\perp+p^\perp+q^\perp)
\eea
where the dilaton operator is taken at the $(0,0)$-picture:
\bea
V_\varphi(k)=\varphi(k)(\partial{X^m}+i(k\psi)\psi^m)
(\bar\partial{X^n}+i(k{\bar\psi}){\bar\psi}^n)
(\eta_{mn}-k_m{\bar{k}}_n-k_n{\bar{k}}_m)
\nonumber \\
k^2={\bar{k}}^2=0
\nonumber \\
(k{\bar{k}})=1
\eea

The crucial point is that
it is only the X-part
of the dilaton operator that contributes to these correlators;
the $\psi$-part vanishes since at least one of the NS fermions
in the dilaton operator, $(q^\perp\psi)$
always has a polarization orthogonal to $\psi's$ in the brane-like
operators and therefore has no partners to be contracted with.
As a result, the relative normalization of this correlator
is one half of those of the dilaton with the usual
perturbative superstring vertices.
As a result, the overal dilaton coupling must be proportional
to $e^{-\varphi}$ and the effective action in the presence of the dilaton
is given by:
\be
S_{eff}(\lambda)=
\int{d^4x}e^{-\varphi}
\sqrt{det(\eta_{ab}+\partial_a\lambda_t\partial_b\lambda^t)}
\ee
We see that the six $\lambda^t$-fields, emerging from
the 6-form H(k) upon imposing the BRST constraints (9)  determine
the location of the D3-brane in the space-time, as has been already
mentioned above. The 4-dimensional momentum $k$ corresponds
to the worldvolume degrees of freedom in the DBI action.

It is remarkable that the effective action (39) derived
from  the closed string sector that we have just computed,
  has an open string D brane-like dilaton dependence.
Below we will give
some heuristic arguments showing that
 such an open-closed string transmutation
is closely related to the logarithmic properties
 of the brane-like states.
Namely, it has been observed previously that closed string
 brane-like states (8), integrated over their four-dimensional momenta,
 constitute a pair of logarithmic operators.
This means that
 the worldsheet logarithmic singularities  will appear in
the OPE of two such states, after the appropriate momentum integration.
Therefore the insertion of two
integrated closed string brane-like vertices produces one logarithmic
branch point on the worldsheet.
Appearance of the branch point  means that two brane-like insertions
cut a hole on the worldsheet which cannot be removed by any metric
redefinition. Such a logarithmic hole effectively changes the
Euler character of the  worldsheet by 1.
But the dilaton dependence  of the effective action is determined
from the tree-point correlator of the dilaton with two brane-like
insertions.
As a result one obtains an open string
 dilaton dependence from closed string
scattering amplitudes.
 Geometrically, the mechanism of the "wrong"
dilaton coupling that we observed,
 is quite analogous to
the anomalous dilaton dependence of the
RR-fields, which again is determined from
 the three-point function of the
dilaton with two RR-vertices.
The difference, however, is that in the case
of the RR-dilaton interaction each of the
RR-insertions pins a separate worldsheet hole,
as each of the RR-vertices changes boundary
conditions for fermions, creating two separate cuts.
As a result, in the RR case the effective Euler character
is changed by two units, contrary to the brane-like case
where it takes at least a couple of vertices
to create a cut.

\section{Open String Brane-like Vertices and D-brane wavefunctions}
So far we were considering  the brane-like states in the closed string
sector, showing that  in the low energy limit their correlations
leads to the "gravitational" part of the DBI action, involving
the induced gravity in the worldvolume. Now
it is important to understand the role of
the brane-like states in the open-string sector.
Below we'll try to show that this role is that
the open string
brane-like states
account for the terms with the Ramond-Ramond charges carried by
D-branes.
Namely, we shall try to show that the open string vertices (1)
can be understood as creation operators for the RR charge sources.
 The full action for D-branes
 also contains the terms involving the coupling of the
worldvolume form to the Ramond-Ramond potential.
It is well-known that D-branes can be understood as closed string
solitons, carrying the RR-charges.
We are going to show  that the vertices (1) also can be regarded
as quantum operators charged with  the RR gauge fields.
To clarify
the relation of the operators (1)
to the standard interpretation of D-branes as the RR
charge sources,
one can think of the following simple analogy with the
QED. One can describe an electron as a classical object with a certain
charge density. In QED, however, the electron described in terms of a
quantum wavefunction $\psi$, giving rise to creation
operators in the second quantized formalism.
Then the fact of the electron carrying the $U(1)$
charge follows from its interaction term with the gauge field in the QED
action, given by $\sim\psi{A}\psi$ where $A$ is the gauge potential.
Therefore in order to show the relation of the vertices (1) to the
Ramond-Ramond charges one has to consider their interaction
with the RR vertices on a disc, showing that
relevant terms in the effective action have the structure
$\sim{dH^{(5)}}{dH^{(5)}}A_{RR}$, where  $H^{(5)}$ is the
space-time five-form of the $V_5$ brane-like  vertex .
Note that the role of the wavefunction must be played
by the gauge-invariant $dH$ field since, due to the BRST conditions
(6)
 the H five-form is defined
up to a gauge transformation by the derivative of a four-form.
In other words, one needs to show that
in the effective action the $H^{(5)}$-field
couples to the RR gauge potential, rather than the RR field strength.
To show this we have to calculate the disc correlator of a RR
vertex operator with two five-forms or the two-forms,
 inserted on the disc boundary.
To insure the correct ghost number balance
(the sum of left and right ghost numbers must be equal to $-2$
on the disc)
one has to consider correlators
\be
A_{5-5-RR}=<V_5^{(-3)}(k;\tau_1)V_5^{(+1)}(p;\tau_2)V_{RR}^{(+1/2,-1/2)}
(q;z,\bar{z})>
\ee
or
\be
A_{2-2-RR}=<V_2^{(-2)}(k;\tau_1)
V_2^{(0)}(p;\tau_2)V_{RR}^{(1/2,-1/2)}
(q;z,\bar{z})>
\ee
i.e. the Ramond-Ramond vertex operator must be taken at
the mixed $(+{1\over2},-{1\over2})$-picture.
Here and elsewhere we shall consider the case of
Neumann boundary conditions on the disc.
The expression for the RR vertex in this picture is given by
\be
V_{RR}^{+1/2,-1/2}(q)
=e^{{1\over2}\phi-{1\over2}{\bar\phi}}\Sigma_\alpha{\bar\Sigma}_\beta
(\partial{X_m}+{i\over4}(q\psi)\psi_m)
{(\gamma^m\gamma^{m_1...m_p})}_{\alpha\beta}{F^{RR}_{m_1...m_p}}(q)
\ee
where $F^{RR}$ is the Ramond-Ramond p-form field strength.
Now, since expressions for the 5-form brane-like states at any
of the two pictures do not contain any dependence on
the derivatives of $X$ or
the momentum (except for the exponent ${e^{ikX}}$),
and the RR-vertex $V_{RR}^{+1/2,-1/2}$
is linear in $\partial{X}$ and $q$, it is clear that the
the overall  amplitude is linear in momentum as well, i.e. it has the
structure
$$S_{HHF}\sim{qHHF_{RR}}\sim{dH H F_{RR}}\sim{dH dH A_{RR}}$$
i.e. it has precisely the form we are looking for;
here $A_{RR}$ stands for the RR gauge potential.

In general case, the expression for this cubic term
involving the higher spin gauge fields,
may be complicated and involve various nontrivial
ways of contraction of indices
 to insure the overall gradient invariance.
The latter is guaranteed by the BRST conditions on $H^{(5)}$.
One only has to insure that the overall contribution is nonzero and
is not reduced to just a topological Chern-Simons term.
At this point we do not yet have a complete classification of all
the RR-charges coupling to the vertices (1). For instance,
it is not yet clear if each of the vertices (1)
couples to a particular RR-form, or to several forms at the same time.
To clarify this point, one has to compute all the set of three-point
disc correlators of the vertices (1) with all the RR-forms which
has not yet been done, as in general these correlators involve rather
lengthy combinations of the cubic terms.
This computation is currently in progress; we hope to
presente it
soon in our future paper.
In this paper we  consider only one precise example of
such an interaction of the vertices (1) to the RR-fields,
demonstrating the coupling of the brane-like forms to the Ramond-Ramond
charges.
Namely, we consider the correlator of the RR five-form
with two brane-like two-form insertions on the disc.
Note that the c-ghost term in the expression (1) for the
$two-form$ at picture zero does not contribute to
the correlator in this case and can be neglected.
Using the expression (1) for the two-forms at pictures $-2$ and 0
and for the RR 5-form at picture $+1/2,-1/2$ we obtain after simple
calculation:
\bea
A_{2-2-RR}=
<V_2^{(-2)}(p)V_2^{(0)}(k)V_{RR-5}^{+1/2,-1/2}(q)>
\nonumber \\
{\sim}Tr(\gamma^m\gamma^{m_1...m_5}\gamma^{n_1n_2}\gamma^{n_3n_4})
q_mF^{RR}_{m_1...m_5}(q)H_{n_1n_2}(p)H_{n_3n_4}(k)
\eea
Evaluating the gamma-matrix trace and making Fourier transform
we easily find corresponding terms in the effective
action:
\bea
S_{2-2-RR}\sim\int{d^{10}X}H_{m_1m_2}(dH)_{m_3m_4m_5}
F_{RR}^{m_1...m_5}+
H\wedge{dH}\wedge{RR}
\eea
The second term is just the topological CS-term while the first one
reflects the fact that the $H$ 3-form field is a quantum
wavefunction that carries the RR charge of a $D3$-brane.
 The gradient invariance of this term
can be easily proven:  there is a manifest  invariance under
the gauge transformations of the Ramond-Ramond field while
the invariance under the gauge transformations of the two-form:
$H^{(2)}\rightarrow{H^{(2)}}+d\Lambda^{(1)}$
can be easily shown by partial integration and using the
 Maxwell's conditions on $F_{RR}$:
$div(F)=\partial_{m_1}{{F^{RR}}_{m_1...m_5}}=0$
which are the BRST constraints for the Ramond-Ramond vertex operator.
After partial integration we can write the first term of (44) as
\be
S_{2-2-RR}\sim\int{d^{10}X}{{\partial_m}H_{mn}}(dH)_{pqr}A_{RR}^{npqr}
\ee
>From the structure of this amplitude we conclude that
indeed the two-form state is the carrier of the 5-form
of the RR-charge; this means that the
three-form brane-like field ${dH}_{mnp}$ constitute a wavefunction for the
D3-brane.
One needs, however, to consider  the correlations of the
two-forms with other Ramond-Ramond fields in order to
point out the full set of the RR-charges generated by the
open string  two-forms
The main conclusion from this computation is that
the brane-like states can be regarded as creation operators for the
D-branes.
\section{Conclusion}
We have demonstrated that the open and close string beane-like vertex
operators describe the dynamics of D-branes in the second quantized
formalism. Closed string vertices induce the gravity in the
D-brane worldvolume while the open string operators account for the
coupling with the RR gauge fields.  It is remarkable that the
DBI action with open string dilaton coupling
appears as a result of the closed string computation.
We argue that this anomalous dilaton coupling is related to the
logarithmic properties of the closed string brane-like states,
such as the creation of the logarithmic branch point on the worldsheet by
the pair of brane-like vertex operators. The logarithmic
behavior of  vertex operators in string theory occurs when their
wavefunction (space-time field)
satisfies the s.c. "logarithmicity criterium",
namely,  in the on-shell limit
it should asymptotically  behave as $H(k){\sim}k^{-N}$ with $N > 6$.
The asymptotic behavior is determined from the
worldsheet beta-function equation involving the corresponding vertices.
Unlike the usual vertex operators, the  brane-like states satisfy
the above criterium.
In our next paper, currently in
progress, we shall provide the detailed analysis of the LCFT properties of
the brane like states,  to peculiarities of the worldsheet RG flow
involving  vertex operators with the ghost-matter mixing.
There are many unanswered questions about the relation of the
ghost-matter mixing and of the brane-like states to non-perturbative
dynamics of D-branes and M-theory; there is also a multitude
of implications for future work in this direction.
One particular problem is giving the full classification
of the RR charges generated by the open string brane-like operators.
Our hope is that  the ghost-matter mixing principle,
apparently allowing us to consider the D-branes in the operator
 formalism and to develop an alternative approach to
non-perturbative brane dynamics, is not confined to string theory,
but also to gauge theories.
It is possible that the ghost-matter mixing
may be a universal phenomenon in various physical theories,
containing  crucial information about
their non-perturbative dynamics.

\vskip1cm

\section{Acknowledgments:}
We would like to thank N.Nekrasov
and A.M.Polyakov
for useful discussions.
I.K. is supported in part by PPARC
rolling grant PPA/G/O/1998/00567 and  EC TMR grants
HPRN-CT-2000-00152 and  HRRN-CT-2000-00148.
D.P. acknowledges the support of the Academy of Finland
under the Project no. 54023 and the hospitality of
Institut des Hautes Etudes Scientifiques (IHES) in
Bures-sur-Yvette where part of this work has been done.

\end{document}